\documentclass{article}
\usepackage{spconf,amsmath,graphicx}
\usepackage{booktabs}
\usepackage{color}
\usepackage{comment}
\usepackage{scalerel}
\usepackage{upgreek}
\usepackage{bbm}
\usepackage{amssymb}
\usepackage{multirow}
\usepackage{csquotes}

\usepackage{xcolor}
\let\lctau\tau 
\renewcommand{\tau}{\scalerel*{\lctau}{X}}
\title{Positive and negative sampling strategies \\ for self-supervised learning on audio-video data}
\name{Shanshan Wang, Soumya Tripathy, Toni Heittola, Annamaria Mesaros\thanks{This work was supported in part by Academy of Finland grant 332063 ``Teaching machines to listen". The authors wish to thank CSC-IT Centre of Science Ltd., Finland,  for providing computational resources.}}
\address{Signal Processing Research Centre\\ 
Tampere University,
Tampere, Finland
}

\begin{document}
\ninept
\maketitle
\begin{abstract}
In Self-Supervised Learning (SSL), Audio-Visual Correspondence (AVC) is a popular task to learn deep audio and video features from large unlabeled datasets. The key step in AVC is to randomly sample audio and video clips from the dataset and learn to minimize the feature distance between the positive pairs (corresponding audio-video pair) while maximizing the distance between the negative pairs (non-corresponding audio-video pairs).  The learnt features are shown to be effective on various downstream tasks. However, these methods achieve subpar
performance when the size of the dataset is rather small. In this paper, we investigate the effect of utilizing class label information in the AVC feature learning task. We modified various positive and negative data sampling techniques of SSL based on class label information to investigate the effect on the feature quality. We propose a new sampling approach which we call soft-positive sampling, where the positive pair for one audio sample is not from the exact corresponding video, but from a video of the same class. Experimental results suggest that when the dataset size is small in SSL setup, features learnt through the soft-positive sampling method significantly outperform those from the traditional SSL sampling approaches. This trend holds in both in-domain and out-of-domain downstream tasks, and even outperforms supervised classification. Finally, experiments show that class label information can easily be obtained using a publicly available classifier network and then can be used to boost the SSL performance without adding extra data-annotation burden.

\end{abstract}


\section{Introduction}
\label{sec:intro}
Self-supervised learning has been a very popular algorithm to learn feature representations \cite{chen2020simple, fonseca2021unsupervised} from a large
number of unlabeled dataset. For robust feature learning in SSL, normally a pretext task is designed \cite{morgado2020learning} to learn feature representation through contrasting positive
pairs against negative pairs. A contrastive loss such as InfoNCE \cite{oord2018representation}, used in the feature learning, pulls the features of positive pairs together and pushes the negative
pairs apart \cite{wang2022self}. Positive pairs are normally formed by two correlated views from the same example, e.g., two
augmented images from the same image \cite{chen2020simple}; two augmented spectrograms of the same audio \cite{fonseca2021unsupervised}; or audio-video pairs from
the same video clip with temporal synchronization between them \cite{korbar2018cooperative, owens2018audio}. Traditionally, negative pairs do not need to be
chosen in any specific way; instead, within a batch, everything else than the positive pair can be a negative pair,
e.g., augmented views of a different image \cite{chen2020simple}, augmented spectrograms of a different audio \cite{fonseca2021unsupervised, wang2023self}, or audio-video pairs
where the audio and the video come from different video clips \cite{morgado2020learning, wang2022self}. This common way of choosing negative pairs in SSL is called random sampling as the negative pairs are formed from randomly chosen items in the dataset. After the feature learning step, these features are
applied to downstream tasks (different from pretext tasks).

However, researchers have found out that better negative sampling can help in learning better feature representations in SSL. While positive pairs
are explicitly available, the vast numbers of combinations possible for forming negative pairs have driven researchers to spend significant effort in finding the best way to select negative pairs \cite{xie2023negative, HAFIDI2022108310}.
Most of these works have found out that using negative pairs that are hard to discriminate by the models, contributes significantly to the success of SSL than the negative pairs that are easily discriminated.
For example, Xie et al. \cite{xie2023negative} proposed a cross-modality score based semi-hard negatives selection approach to improve the performance in a text-to-audio and audio-to-text retrieval system. Moreover, in the
ablation studies of \cite{morgado2020learning}, authors pointed out that the combination of easy-negative sampling (sampled from different video instances) and hard-negative sampling (sampled from different viewpoints, but from the same video frame)
yields better performance than using easy-negative alone. As in SSL, the training data is not labeled, the hardness of the negative pair selections mostly come from random sampling, some score based methods \cite{xie2023negative} or simple intuition \cite{morgado2020learning} which is prone to mistakes. This can lead to mistakenly labelling positive pairs as negatives and the contrastive loss can push away these similar features which is counter-intuitive. For example, in Figure.1 (a) using the random sampling strategy can lead to push dog features away from another dog image. Without the class labels, there is no explicit way to avoid such scenarios in SSL. If the dataset is small the effect of these mistakes can be severe and harm the overall quality of the learnt features.

\begin{figure*}[!t]
\centering
\includegraphics[width=.89 \textwidth]{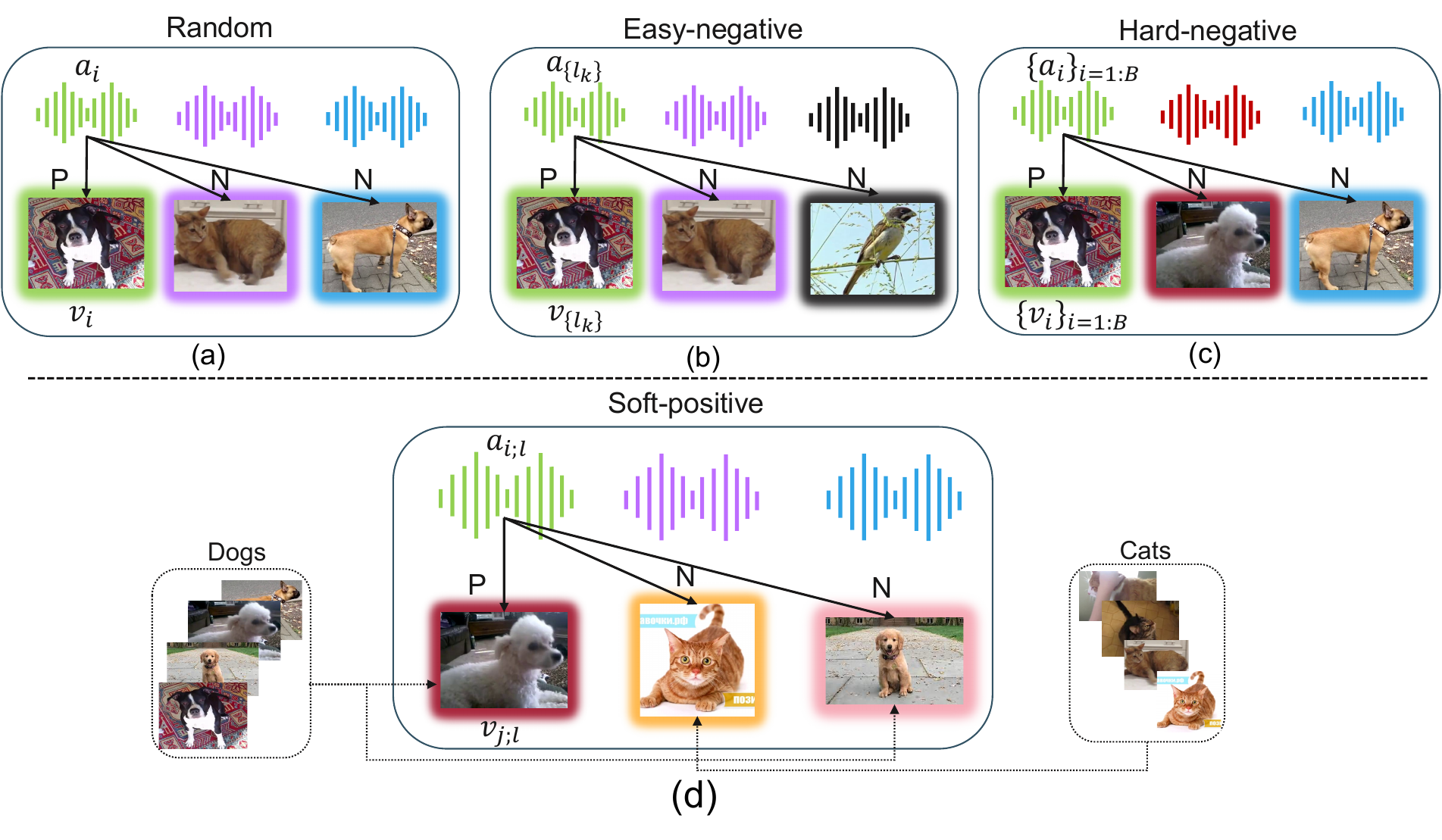}
\vspace{-10pt}
\caption{Illustration of various positive (P) and negative (N) sampling methods: (a) random sampling; (b) easy-negative sampling; (c) hard-negative sampling;  (d) proposed soft-positive sampling. }
\label{fig: diagrams}
\vspace{-6pt}
\end{figure*}

In this work, we investigate different sampling strategies using a class-labeled dataset for audio-visual correspondence  based SSL learning. Based on the labels, we can explicitly control the way pairs are formed, even though we are performing SSL. We want to answer the following research questions. Firstly, for the random negative sampling method, are the learnt features harmed given the fact that negative pairs might contain positive samples? Would the learnt features be better if we make sure within each batch that samples are drawn from truly negative pairs, i.e. different classes? Secondly, what if, within each batch, we draw samples from the same class only, i.e., we force learning only with hard negatives? Thirdly, while choosing the positive pairs is straightforward and obvious, is it possible, and, if so, beneficial to select positive pairs in a different way? 

Finally, we propose a novel sampling approach for choosing the positive pairs called soft-positive sampling. This is contrary to the common belief that negative sampling is the most crucial part of SSL. Unlike the default way of pairing the audio and images of the corresponding video clip, with soft-positive sampling we pair the audio with images from the video clips of the same class. Experimentally, using soft-positive pairs improves the downstream performance significantly, especially on smaller datasets, than other sampling strategies.



\section{Method}
\label{sec:method}

For SSL using audio-visual data, the most common pretext task is audio-visual
correspondence. In simpler terms, the network learns robust features by learning to match the corresponding audio and
video pairs in a data batch. Such networks are trained using the InfoNCE loss which minimizes feature distance
between corresponding audio-video pairs (positive pairs) and maximizes the distance for non corresponding audio-video
pairs (negative pairs) in a sampled mini-batch of audio-video data. These mini-batches are constructed by sampling the
audio and videos from the dataset.

For SSL we consider a dataset $\{a_i, v_i\}_{i=1:N}$ with total N number of audio samples $a_i$ and video samples $v_i$ across $C$ class. Figure \ref{fig: diagrams} (a-d) illustrates the different sampling methods examined throughout this paper. We use color to indicate
the exact correspondence between audio and video, and illustrate the cases using a batch size of $B=3$.  Figure \ref{fig:
diagrams} (a) shows the random sampling method where we randomly pick $B$ number of $v_i$ and their corresponding $a_i$. The audio-video pair is a positive sample if the audio and the video are extracted from the same video clip i.e. $\{a_i, v_j\}$ where $(i=j)$ and temporally aligned, while any other combination of pairing $\{a_i, v_j\}$ with $(i \neq j)$ is a negative pair. The obvious demerit of such a technique is that multiple samples in a mini-batch may belong to the same class label, in this illustration \textit{dog}.

To understand the effect of different sampling methods for negative pairs, we design two carefully controlled sampling
techniques: (1) easy-negative, illustrated in Fig. \ref{fig: diagrams} (b), where we first randomly sample $B$ classes
from $\{l_k\}_{(k=1:C)}$ and from each class $l_k$ we draw one $v_{\{l_k\}}$ and its corresponding $a_{\{l_k\}}$. This makes sure that the negative samples
are drawn from different classes within each batch and can be relatively easy for the network to discriminate;
and (2) hard-negative, illustrated in Fig. \ref{fig: diagrams} (c), where we first randomly sample a class from
$\{l_k\}_{k=1:C}$ and then randomly draw $B$ number of $\{v_i\}_{i=1:B}$ and their corresponding $\{a_i\}_{i=1:B}$ to ensure that the  negative samples come always from the
same class (in this case \textit{dog}) but from different instances than the positive pair, which is a very hard
condition for the network to learn.

All the sampling methods described above focus on the negative pairs selection. We propose a new sampling method called
\textit{soft-positive sampling}, illustrated in Fig. \ref{fig: diagrams} (d). The core idea of this method is to first
randomly pick $B$ number of $a_i$ and their corresponding label $l$. Instead of picking the corresponding videos, we pick $v_j$ where $i \neq j$ but they correspond to same labels $l$. This creates positive pairs between $a_{i;l}$ and $v_{j;l}$ where $i \neq j$ but $a_i$ and $v_j$ belongs to same class $l$.  In this way we effectively use a hard-negative as a
soft-positive. As shown in Fig. \ref{fig: diagrams} (d), the audio exactly matching the dog in green is paired as a
positive pair with a different dog image (in this case, in red). Similarly, the audio of cat in purple would form a
positive pair with the image of the cat in yellow, instead of the exact match. This way of designing soft-positive pairs
enforces variability for the data. More importantly, unlike the traditional way of forming the positive pair which
enforces the network not only to learn the correspondence of audio and the video but also ties it to the audio-visual
temporal synchronization, soft-positive sampling allows the freedom for the network to reason the correspondence and
correlations between the audio and video content in a broader manner.

\section{Experimental setup}
\label{sec:setup}

\subsection{Datasets}
\label{sec:data}
Three datasets are used throughout all the experiments: VGGSound \cite{9053174}, having more than $210$k videos of $10$ seconds each, and  including $309$ audio classes; TAU Urban Audio-Visual Scenes 2021 \cite{wang2021curated}, having $12,292$ videos of $10$ seconds each, and including $10$ audio classes; and FSDnoisy18k \cite{fonseca2019learning}, containing $42.5$ hours of audio across $20$ sound event classes. VGGSound is used to train the pretext task of audio-visual correspondence, and for the in-domain downstream task, while TAU Urban Audio-Visual Scenes 2021 and FSDnoisy18k datasets are used to test out-of-domain downstream tasks. Note that the reason why we choose VGGSound for the pretext task training is that it is a curated dataset which ensures the audio-visual correspondence and has rather clean audio class labels compared to AudioSet \cite{7952261}. With the clean label information, we can investigate the effect of the different sampling methods described above. 

To simulate the effect of a smaller dataset for SSL and to better utilize the class labels, we select a fixed balanced subset of VGGSound dataset. This subset consists of all $309$ classes, and has $123$ samples in each class (the size of the smallest class in the dataset), summing up to 38,007 video clips.  Easy-negative sampling requires that within each batch, samples must come from different classes, so we set the batch size to $309$, which is equal to the total number of audio classes.

\begin{table*}[t]
    \centering
    \begin{tabular}{l|c|c|c|c|c|c}
        \toprule
         & \multicolumn{2}{|c|}{In-domain test} & \multicolumn{2}{|c}{Out-of-domain test on TAU} &\multicolumn{2}{|c}{Out-of-domain test on FSD} \\
         Sampling method& linear evaluation & finetune & linear evaluation & finetune & linear evaluation & finetune\\
        \midrule
        Random & 13.8\% & 31.3\% & 59.6\% & 61.7\% & 44.9\% &59.8\%\\
        \midrule
        Easy-negative & 13.1\% & 30.9\% & 57.7\% & 62.4\% &48.2\%* & 61.1\% \\
       \midrule
        Hard-negative& 6.9\% & 30.8\% & 58.7\% & 62.3\% & 36.6\% & 56.8\%\\
        \midrule
        Soft-positive (50\%) & 19.7\%* & 31.3\% & 59.4\% & 62.9\% &\textbf{54.4\%} & 65.8\%* \\
        \midrule
        \textbf{Soft-positive (proposed)} & \textbf{20.5\%} & \textbf{32.3\%} & 59.8\% & 64.6\%* & 52.9\%* & \textbf{68.1\%} \\
        \midrule 
        \textbf{PL-soft-positive} & 16.2\%* & 30.4\% & 58.2\% & 62.8\% & 48.9\%* & 66.1\%*\\
        \midrule
        Supervised & \multicolumn{2}{|c|} {31.7\%}  & \multicolumn{2}{|c} {61.6\%} &\multicolumn{2}{|c} {55.2\%}\\
        \bottomrule

    \end{tabular}
    \caption{In-domain and out-of-domain classification accuracy for different sampling methods trained on the balanced subset of VGGSound. 
    Bold numbers indicate the best among all sampling methods; * indicates that the accuracy is significantly better than for the random sampling.}
    \label{tab: results}
\end{table*}

\subsection{Pretext task - AVC }
In the pretext AVC training, we follow the setup from \cite{morgado2020learning}. For each $10$-seconds clip, we randomly select $0.5$ seconds of video and $1$ second of audio covering the selected video frames. Video frame rate is set to $16$, meaning that $8$ random frames are extracted for each video clip. Frames are further resized into $(112,112)$, and a $50\%$ probability of frames being flipped from left to right is used as video augmentation during training. The images are normalized and further sent to the video encoder with size of $[B, 3, 8, 112, 112]$. In case of audio, the sampling rate is set to $16$ kHz, fft size is set to $512$, frame rate is set to $100$, and mel bands are set to $64$. The log-mel spectrogram is normalized and then forwarded to the audio encoder with size of $[B, 1, 100, 64]$. 

The video encoder adopts the R2+1D network with 18 layers as \cite{8578773}, and the audio encoder employs a 9-layer 2D Convolutional Neural Network (CNN). For the random, easy-negative and the proposed soft-positive sampling methods, we use a batch size of $309$. For the hard-negative sampling method we set the batch size to $123$, to ensure that within each batch, samples come from the same class only. We train the network using  InfoNCE contrastive loss, Adam optimizer with a learning rate of $0.0001$ and weight decay of $0.00001$. We train all the networks for $100$ epochs, using Pytorch framework \cite{NEURIPS2019_9015}. We use $4$ Tesla A100 GPUs with 40GB memory in each GPU for the each model training. The code is openly available\footnote{https://github.com/shanwangshan/positive-negative-sampling-strategy}. 

\subsection{Downstream audio classification tasks}
We test the effectiveness of the features learnt through different sampling methods by designing in-domain and out-of-domain downstream audio classification tasks. For the downstream tasks, the video encoder is discarded, and only the audio encoder is used, with an additional linear classifier layer whose output neurons are defined according to the task at hand. The training for the downstream task uses only audio. We evaluate the methods using two approaches: linear evaluation and finetune. For linear evaluation, only the last linear classification layer is trained, while for finetune the whole audio encoder is trained using the downstream task data.

\begin{table*}
    \centering
    \begin{tabular}{l|c|c|c|c|c|c|c}
        \toprule
         & \multicolumn{3}{|c|}{In-domain test} & \multicolumn{2}{|c}{Out-of-domain test on TAU}&\multicolumn{2}{|c}{Out-of-domain test on FSD} \\
         Sampling method& linear evaluation & finetune (5\%) & finetune & linear evaluation & finetune & linear evaluation & finetune  \\
        \midrule
        Random & 18.4\% & 19.4\% &32.4\% & 60.4\% & 67.2\%& 54.0\% & 69.1\% \\
        \midrule
        \textbf{Soft-positive } & \textbf{23\%} & \textbf{20.6\%} &32.3\% & 59.8\% & 68.4\% & 56.4\% & 68.4\%\\
        \midrule
        \textbf{PL-soft-positive} & 18.2\% & 19.3\% &32.2\% & 59.4\% & 68.4\% & 54.1\% & 68.8\% \\
        \midrule
        Supervised & \multicolumn{2}{|c|} {31.9\%}  & \multicolumn{2}{|c} {61.6\%} &\multicolumn{2}{|c} {55.2\%}\\
        
        \bottomrule
        
    \end{tabular}
    \caption{In-domain and out-of-domain classification accuracy for different sampling methods trained on the complete VGGSound dataset. 
    Bold numbers indicate the best among all sampling methods; * indicates that the accuracy is significantly better than for the random sampling.}
    \label{tab: results_complete}
    
\end{table*}

\textbf{In-domain tasks}: 
For in-domain audio classification, we randomly split the balanced training subset we created from VGGSound with a ratio of $80\%$ and $20\%$, to use for training and validation, respectively. After the AVC training, we append one linear layer of 309 units to the audio encoder, and train it as a classification layer. For training, we randomly select $1$ second of audio from each clip.  
We train the network using cross entropy loss with a batch size of $64$ for $100$ epochs with validation. The test set of VGGSound includes $14,518$ audio samples across $309$ audio classes. For testing we use the whole $10$ seconds of audio. Classification accuracy is used as the evaluation metric. 

\textbf{Out-of-domain tasks:}
For out-of-domain audio classification, the overall training and evaluation framework are similar to the in-domain task. The only differences are the input data and the number of neurons in the last linear layer. TAU data is split into the same ratio  $80/20$. The test set consists of $3645$ audio samples across $10$ classes. During training, we randomly select 2 seconds of audio, while for testing, we use the whole 10 seconds of audio. The model consists of the audio encoder with one extra linear layer having $10$ output units.  

For FSD dataset, we train the system using only the clean subset of the data, having $1772$ samples and use the same $80/20$ split ratio. The test set includes $947$ samples. The model consists of the audio encoder with one extra linear layer having $20$ output units. During training, $1$s of audio is used, similar as in \cite{fonseca2021unsupervised}, and the whole clip is used in the testing case. 

\subsection{Soft-positive sampling}
To investigate the effect of soft-positive sampling in a more realistic scenario, we design the following experiments: firstly, we utilize the ground truth labels information directly to perform soft-positive sampling. In order to simulate the scenario where there are no ground truth labels available, we propose to use a pre-trained classifier to assign pseudo-labels for datasets. To achieve the pseudo-labelling, we use EfficientNet \cite{tan2020efficientnet} and perform classification on the images of VGGSound dataset. Then, we perform soft-positive sampling based on the pseudo labels provided by the image classifier. For simplicity, we abbreviate this pseudo-label-soft-positive method to \textit{PL-soft-positive}. As an ablation study, we combine soft-positive and exact positive sampling (as in random sampling) with a probability of 0.5 to test the contribution of soft-positive sampling. We denote this study as \textit{Soft-positive (50\%)}. 
Finally, a supervised classification on each of the datasets is performed as a baseline, using the same audio encoder architecture but trained from scratch using the labels.

In order to compare the soft-positive sampling strategy based SSL in a bigger dataset, we perform the same experiments using the complete VGGSound dataset (210k audio-video data with unbalanced classes), for all different sampling methods for SSL. 

\section{Results and discussions}
\label{sec:results}
Audio classification results for the in-domain and out-of-domain tasks are shown in Table \ref{tab: results}  for the systems trained on the small, balanced subset of VGGSound dataset. 
Looking at the first two rows, the results suggest that choosing random sampling or easy-negative sampling does not make a big difference if the downstream task data is large (in-domain, and out-of-domain TAU data). However, if the data used to train for the downstream task is rather small, in this case the FSD clean data with $1772$ samples, the easy-negative sampling shows a clear advantage (48.2\% for easy-negative sampling compared to 44.9\% for random sampling  with the linear evaluation). Based on these results, it seems that at least for the AVC proxy task there is no need for concern that same class samples may appear in the same batch while calculating the contrastive loss. Apparently, this loss has its internal mechanism to place the features reasonably. However, if the downstream task data is rather small in size, then it is beneficial to consider other sampling strategies. 

The features learnt through the hard-negative sampling method are not completely rubbish, even though this approach achieves the lowest classification accuracy compared to other sampling methods for linear evaluation. As seen from Table \ref{tab: results}, the performance gap is filled by fine-tuning the audio encoder during training for the downstream task. 

The proposed soft-positive sampling method outperforms other sampling methods significantly in many test cases: compared to the conventional random sampling method, it achieves around $7\%$ higher in linear evaluation for in-domain tasks. As for out-of-domain tasks, it also shows a consistent advantage: around $3\%$ higher in finetune evaluation for TAU dataset and around $8\%$ higher both in linear evaluation and fine-tune evaluation for FSD dataset. Comparing with the supervised learning, it can be seen that SSL with random sampling and fine-tuning reaches similar performance to supervised learning. However, with the trick of soft-positive sampling in SSL and fine-tuning we achieve better results than with a supervised baseline.
This is also observed from the \textit{soft-positive (50\%)} case: the performance is much higher than random sampling, and almost on par with the soft-positive sampling. This further demonstrates the effectiveness of soft-positive sampling. Moreover, while the PL-soft-positive approach brings better performance than random sampling only for the FSD dataset, the results suggest that by assigning pseudo labels using pre-trained classifier brings benefit (or at least do not harm) the feature representation compared to the random sampling. 

Table \ref{tab: results_complete} shows the in-domain and out-of-domain test results for training the system on the complete VGGSound dataset. Results show that soft-positive sampling still holds the advantages for in-domain tasks, while achieving a similar performance for out-of-domain tasks. 
However, results of  PL-soft-positive are no different compared to random sampling. One possible explanation is that the classifier used to assign pseudo labels achieves a low accuracy on the VGGSound images, due to the fact that EfficientNet is trained on ImageNet \cite{ deng2009imagenet} with 1000 classes and there are strong mismatches in terms of labels between ImageNet and VGGSound. The main difference is that ImageNet labels are provided based on the image content while VGGSound classes are labelled based on the audio content.


\section{Conclusions}
We investigated various sampling strategies for choosing negative and positive pairs in self-supervised learning using audio-video data in AVC task. We propose a novel sampling technique, soft-positive sampling, which improves the feature representation significantly when the dataset for pretext task is relatively small and balanced. We also observe that random sampling does not harm the feature representation when the downstream task data is large enough. However, when the downstream task data is small, it is sub-optimal to use random sampling and it is beneficial to adopt the soft-positive sampling strategy.

\label{sec:concl}

\newpage
\bibliographystyle{IEEEbib}
\bibliography{references}

\begin{thebibliography}{10}

\bibitem{chen2020simple}
Ting Chen, Simon Kornblith, Mohammad Norouzi, and Geoffrey Hinton,
\newblock ``A simple framework for contrastive learning of visual representations,''
\newblock in {\em International conference on machine learning}. PMLR, 2020, pp. 1597--1607.

\bibitem{fonseca2021unsupervised}
Eduardo Fonseca, Diego Ortego, Kevin McGuinness, Noel~E. O'Connor, and Xavier Serra,
\newblock ``Unsupervised contrastive learning of sound event representations,''
\newblock in {\em 2021 IEEE International Conference on Acoustics, Speech and Signal Processing (ICASSP)}. IEEE, 2021.

\bibitem{morgado2020learning}
Pedro Morgado, Yi~Li, and Nuno Nvasconcelos,
\newblock ``Learning representations from audio-visual spatial alignment,''
\newblock {\em Advances in Neural Information Processing Systems}, vol. 33, 2020.

\bibitem{oord2018representation}
Aaron van~den Oord, Yazhe Li, and Oriol Vinyals,
\newblock ``Representation learning with contrastive predictive coding,''
\newblock {\em arXiv preprint arXiv:1807.03748}, 2018.

\bibitem{wang2022self}
Shanshan Wang, Archontis Politis, Annamaria Mesaros, and Tuomas Virtanen,
\newblock ``Self-supervised learning of audio representations from audio-visual data using spatial alignment,''
\newblock {\em IEEE Journal of Selected Topics in Signal Processing}, 2022.

\bibitem{korbar2018cooperative}
Bruno Korbar, Du~Tran, and Lorenzo Torresani,
\newblock ``Cooperative learning of audio and video models from self-supervised synchronization,''
\newblock {\em Advances in Neural Information Processing Systems}, vol. 31, 2018.

\bibitem{owens2018audio}
Andrew Owens and Alexei~A Efros,
\newblock ``Audio-visual scene analysis with self-supervised multisensory features,''
\newblock in {\em Proceedings of the European conference on computer vision (ECCV)}, 2018, pp. 631--648.

\bibitem{wang2023self}
Shanshan Wang, Soumya Tripathy, and Annamaria Mesaros,
\newblock ``Self-supervised learning of audio representations using angular contrastive loss,''
\newblock in {\em ICASSP 2023-2023 IEEE International Conference on Acoustics, Speech and Signal Processing (ICASSP)}. IEEE, 2023, pp. 1--5.

\bibitem{xie2023negative}
Huang Xie, Okko R{\"a}s{\"a}nen, and Tuomas Virtanen,
\newblock ``On negative sampling for contrastive audio-text retrieval,''
\newblock in {\em ICASSP 2023-2023 IEEE International Conference on Acoustics, Speech and Signal Processing (ICASSP)}. IEEE, 2023, pp. 1--5.

\bibitem{HAFIDI2022108310}
Hakim Hafidi, Mounir Ghogho, Philippe Ciblat, and Ananthram Swami,
\newblock ``Negative sampling strategies for contrastive self-supervised learning of graph representations,''
\newblock {\em Signal Processing}, vol. 190, pp. 108310, 2022.

\bibitem{9053174}
Honglie Chen, Weidi Xie, Andrea Vedaldi, and Andrew Zisserman,
\newblock ``Vggsound: A large-scale audio-visual dataset,''
\newblock in {\em ICASSP 2020 - 2020 IEEE International Conference on Acoustics, Speech and Signal Processing (ICASSP)}, 2020, pp. 721--725.

\bibitem{wang2021curated}
Shanshan Wang, Annamaria Mesaros, Toni Heittola, and Tuomas Virtanen,
\newblock ``A curated dataset of urban scenes for audio-visual scene analysis,''
\newblock in {\em ICASSP 2021-2021 IEEE International Conference on Acoustics, Speech and Signal Processing (ICASSP)}. IEEE, 2021, pp. 626--630.

\bibitem{fonseca2019learning}
Eduardo Fonseca, Manoj Plakal, Daniel~PW Ellis, Frederic Font, Xavier Favory, and Xavier Serra,
\newblock ``Learning sound event classifiers from web audio with noisy labels,''
\newblock in {\em ICASSP 2019-2019 IEEE International Conference on Acoustics, Speech and Signal Processing (ICASSP)}. IEEE, 2019, pp. 21--25.

\bibitem{7952261}
Jort~F. Gemmeke, Daniel P.~W. Ellis, Dylan Freedman, Aren Jansen, Wade Lawrence, R.~Channing Moore, Manoj Plakal, and Marvin Ritter,
\newblock ``Audio set: An ontology and human-labeled dataset for audio events,''
\newblock in {\em 2017 IEEE International Conference on Acoustics, Speech and Signal Processing (ICASSP)}, 2017, pp. 776--780.

\bibitem{8578773}
Du~Tran, Heng Wang, Lorenzo Torresani, Jamie Ray, Yann LeCun, and Manohar Paluri,
\newblock ``A closer look at spatiotemporal convolutions for action recognition,''
\newblock in {\em 2018 IEEE/CVF Conference on Computer Vision and Pattern Recognition}, 2018, pp. 6450--6459.

\bibitem{NEURIPS2019_9015}
Adam Paszke, Sam Gross, Francisco Massa, Adam Lerer, James Bradbury, Gregory Chanan, Trevor Killeen, Zeming Lin, Natalia Gimelshein, Luca Antiga, Alban Desmaison, Andreas Kopf, Edward Yang, Zachary DeVito, Martin Raison, Alykhan Tejani, Sasank Chilamkurthy, Benoit Steiner, Lu~Fang, Junjie Bai, and Soumith Chintala,
\newblock ``Pytorch: An imperative style, high-performance deep learning library,''
\newblock in {\em Advances in Neural Information Processing Systems 32}, H.~Wallach, H.~Larochelle, A.~Beygelzimer, F.~d\textquotesingle Alch\'{e}-Buc, E.~Fox, and R.~Garnett, Eds., pp. 8024--8035. Curran Associates, Inc., 2019.

\bibitem{tan2020efficientnet}
Mingxing Tan and Quoc~V. Le,
\newblock ``Efficientnet: Rethinking model scaling for convolutional neural networks,'' 2020.

\bibitem{deng2009imagenet}
Jia Deng, Wei Dong, Richard Socher, Li-Jia Li, Kai Li, and Li~Fei-Fei,
\newblock ``Imagenet: A large-scale hierarchical image database,''
\newblock in {\em 2009 IEEE conference on computer vision and pattern recognition}. Ieee, 2009, pp. 248--255.

\end{thebibliography}

\end{document}